\documentstyle[12pt,epsf]{article}
\newcommand{\be}{\begin{equation}}
\newcommand{\ee}{\end{equation}}
\newcommand{\bea}{\begin{eqnarray}}
\newcommand{\eea}{\end{eqnarray}}
\newcommand{\Tr}{\mathop{\rm Tr}\nolimits}

\newcommand{\Complex}{
       \mbox{C \hspace{-1.16em} \raisebox{-0.018em}{\sf l}}\;}

\newcommand{\half}{{\textstyle \frac{1}{2} }}
\title{A Solvable Model for Intersecting Loops}
\author{Bernard Nienhuis\thanks{e-mail: \tt nienhuis@phys.uva.nl}\\
Ronald Rietman\thanks{e-mail: \tt rietman@phys.uva.nl}\\
Instituut voor Theoretische Fysica\\ Valckenierstraat 65\\
1018 XE Amsterdam\\ The Netherlands}
\date{ITFA 92--35, hep-th/9301012, December 1992}

\begin{document}

\maketitle
\begin{abstract} We show that some models with non-local (and
non-localizable) interactions have a property, called quasi-locality, which
allows for the definition of a transfer matrix. We give the Yang-Baxter
equation as a sufficient condition for the existence of a family of
commuting transfer matrices and solve them for a loop model with
intersections. This solvable model is then analyzed in some detail and its
applications to a Lorentz gas are briefly discussed.
\end{abstract}
\epsfverbosetrue
In the field of solvable models in statistical mechanics it is customary
to restrict oneself to models with local interactions. This restriction
makes it hard to treat models of extended objects, such as linear or
branched polymers, directly. The usual approach is to consider the graphical
expansion of a bona-fide model with local interactions, such as the Potts
model or an $O(n)$ model, and derive properties of the graphical model from
the local model. This is not always possible. In this paper we present a
model, the intersecting loop model, for which there is no local
representation as far as we know.

We consider a dense gas of loops on a square lattice.
In contrast to
the loop gas that one obtains in the graphical expansion
of the Potts model or a RSOS model, our loops are allowed to
intersect themselves and each other.

On every site of a square lattice we randomly put a mirror along one of the
diagonals, or leave it empty. The mirrors deflect photons that run along the
edges of the lattice. If we impose toroidal boundary conditions every photon
follows a closed orbit.
Obviously every edge belongs to precisely one path.
In this paper we are
mainly interested in the statistics of the paths, i.e.\ given the
probabilities that a site is empty or that it carries a mirror
oriented along the NE-SW or the NW-SE
diagonal we want to calculate
quantities such as the density of paths and their length distribution.
The kinetic properties of this Lorentz gas, such as its anomalous diffusive
behaviour, have been sudied in~\cite{cohen}.

We take the probabilities
$P_i(\hbox{\rm NE-SW})$, $P_i(\hbox{\rm NW-SE})$ and $P_i({\rm empty})$ that
the
mirror at site $i$ is oriented along the NE-SW diagonal, the NW-SE diagonal
or absent independent of $i$. For convenience we will mostly work with the
Boltzmann weights (unnormalized probabilities), denoted respectively by $a$,
$b$ and $c$. Furthermore we give every loop a fugacity $q$. The oject of our
study is the partition sum (or generating function)
\be
Z = \sum_{\rm mirror\atop configurations} a^{\#\hbox{\scriptsize NE-SW}}
b^{\#\hbox{\scriptsize NW-SE}}
c^{\#\rm empty} q^{\#\rm paths} .\ee

When $c=0$ this model can be mapped onto a 6 vertex model and
it is the graphical representation of the self-dual $q^2$-states Potts model.
However, if $c\ne 0$ no mapping to a model with local interactions is known,
unless $q$ is an integer $>1$, hence a different approach is needed.

Our approach is based on the observation that, although the Boltzmann
weight is non-local, it is possible to obtain the Boltzmann weight for a
configuration on a large patch of the lattice by glueing together smaller
patches. This property may be called {\em quasi-locality}.

Let $A$ and $B$ be neighbouring patches of the lattice.
To every mirror configuration on $A$ we assign a Boltzmann weight $W(A)$
(a factor $a$, $b$ or $c$ for each site and a factor $q$ for each closed
loop in $A$) and a {\em connectivity}, indicating which pairs of edges on
the boundary $\partial A$ of $A$ lie on the same photon path.
We do the same for $B$.
Now we can calculate the Boltzmann weight and connectivity of the total
patch $AB$. The connectivity of $\partial AB$ is induced by the connectivities
of $\partial A$ and $\partial B$
(glue the paths together on the common boundary $\partial A \cap \partial B$
of $A$ and $B$), and the Boltzmann weight is
\be
W(AB) = W(A) W(B) q^{\#(\partial A\cap \partial B)},
\ee
where $\#(\partial A\cap \partial B)$ is the number of paths closed
by glueing $A$ and $B$ together.

This property makes it possible to define a transfer `matrix', acting in the
space of connectivities of the boundaries of a cylinder.
It is defined via
\be   \label{deftm}
Z_{M+1}(\alpha) = \sum_{\beta} T_{\alpha\,\beta} Z_M(\beta).
\ee
Here $\alpha$ and $\beta$ denote connectivities of the boundaries of a
cylinder (the circumference $N$ is implicit in the equations) and
$Z_M(\alpha)$ is defined as the sum over all mirror configurations
that lead to the connectivity $\alpha$
of the Boltzmann weight for mirrors and closed loops on a cylinder of height
$M$. This definition of a transfer matrix is similar to the one for the
$Q$-states Potts model in~\cite{blotnight}.

Since
\be
Z_0(\alpha) = \delta_{\alpha\,{\sf id}},
\ee
where $\sf id$ is the trivial connectivity, in which every edge in the top
row is connected to the edge in the bottom row straight below it,
\be
Z_M(\alpha) = T^M_{\alpha\,\sf id}.
\ee
The toroidal partition function is obtained by closing the cylinder. This
yields a factor $q$ for every newly formed loop; for each connectivity
$\alpha$ we denote the number of loops by $\#(\alpha)$.
Finally we sum over $\alpha$:
\be
Z_M = \sum_{\alpha} q^{\#(\alpha)} T^M_{\alpha\,\sf id}.
\ee

We now rephrase these observations in a slightly more
formal language.
The connectivities of the upper and lower boundaries can be taken as the
basis of an algebra. Addition is defined formally and
multiplication is defined by glueing the connectivities together, again
giving a factor $q$ for each loop that is formed.
Since there are $(2N-1)!!$ connectivities, the algebra of connectivities
$B_N(q)$ (a.k.a.\ the Brauer algebra, since it was discovered by
Brauer~\cite{brauer} as the
algebra of invariants of $SO(q)$ for integer $q$) is $(2N-1)!!$ dimensional.

The connectivities are most naturally represented by pictures.
Each picture consists of two rows of $N$ dots each above each other,
representing the edges on the upper and lower boundary, and lines connecting
pairs of dots, indicating which edges lie on the same path. To calculate the
product $\alpha\beta$, identify the bottom row of $\alpha$ with the top row
of $\beta$, remove the dots on this middle row, straighten out the lines and
replace every closed loop by a factor $q$.

Instead of considering a connectivity and its partition function separately,
we can view them as an element of the Brauer algebra:
\be
Z_M(\alpha) \mapsto Z_M(\alpha) \cdot \alpha.
\ee
If we now define
\be
{\sf Z}_M = \sum_{\alpha} Z_M(\alpha)\cdot\alpha
\ee
the transfer matrix, defined in eq.~(\ref{deftm}), becomes an element of
the Brauer algebra,
\be
{\sf Z}_{M+1} = {\sf T} {\sf Z}_M.
\ee
The relation with the `components' $T_{\alpha\,\beta}$ given before is
\be
{\sf T}\beta = \sum_{\alpha} T_{\alpha\,\beta} \cdot\alpha.
\ee
We denote a configuration of a row of mirrors by $\vec{m}=
(m_1,m_2,\ldots,m_N)$, where $m_i=1$, $-1$ or $0$ according to whether site
$i$ has a NE-SW mirror, a NW-SE mirror or is empty.
To every $\vec{m}$  corresponds a connectivity
$\alpha(\vec{m})$, so the transfer matrix can be written
as
\be
{\sf T} = \sum_{\vec{m}} w(\vec{m})\cdot\alpha(\vec{m}),
\ee
where
\be
w(\vec{m}) = q^{\delta({\vec{m},\vec{0}})} a^{\#(m_i=1)} b^{\#(m_i=-1)}
c^{\#(m_i=0)}.
\ee
The transfer matrix is the linear combination of connectivities
\be
{\sf T} = \;\;\vcenter{\epsffile{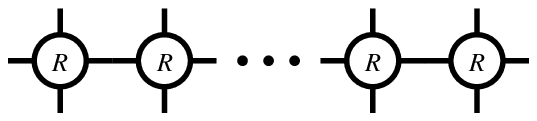}}\;,
\ee
where the leftmost horizontal edge should be connected to the rightmost
horizontal edge and where
\be
\vcenter{\epsffile{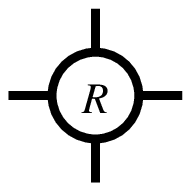}}\; =\; a\;\; \vcenter{\epsffile{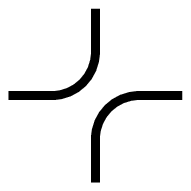}}\;
 + b\;\; \vcenter{\epsffile{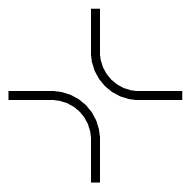}}\;
+ c \;\;\vcenter{\epsffile{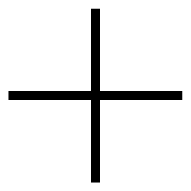}}\;.
\ee

Since ${\sf Z}_0 = {\sf id}$,
\be
{\sf Z}_M = {\sf T}^M.
\ee
The partition function $Z_M$ can now be written as
\be
Z_M = \Tr {\sf Z}_M,
\ee
where the trace is defined as a linear function $B_N(q) \to \Complex$ with
\be \Tr \alpha = q^{\#(\alpha)}. \ee
The trace satisfies
\be \Tr (\alpha\beta) = \Tr(\beta\alpha).\ee
A simple example may clarify the definitions. Consider the algebra $B_4(q)$.
It is 105-dimensional and two connectivities are
\be
\alpha =\;\;\vcenter{\epsffile{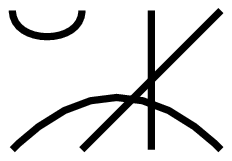}}\;\quad\hbox{and}\quad
\beta = \;\;\vcenter{\epsffile{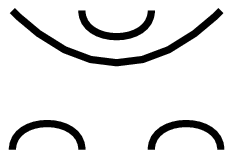}}\;.
\ee
Their products are
\be
\alpha\beta=q\;\;\vcenter{\epsffile{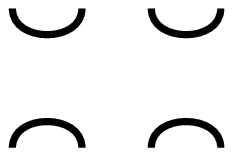}}\;\quad\hbox{and}
\quad\beta\alpha=q\;\;\vcenter{\epsffile{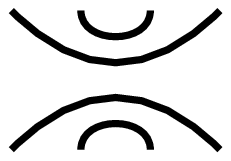}}\;
\ee
and the traces are
\be
\Tr\alpha = q^2\quad \Tr\beta = q \quad \Tr(\alpha\beta) = \Tr(\beta\alpha)
= q^3.
\ee

The usual lattice models with local interactions are called solvable when
their Boltzmann weights satisfy the Yang-Baxter equation. This
equation implies that the transfer matrices constructed with the aid of
these Boltzmann weights commute for arbitrary system sizes.
We can do something similar for the loop model: when the mirror probabilities
$a$, $b$ and $c$ satisfy the Yang-Baxter equation for loop models, the
corresponding transfer matrices ${\sf T}$ commute for arbitrary system sizes.
The Yang-Baxter equation for loop models states that the l.h.s.\ and the
r.h.s.\ of eq.~(\ref{ybepic}) are the same linear combination of
connectivities of the six outer edges.
\be
\vcenter{\epsffile{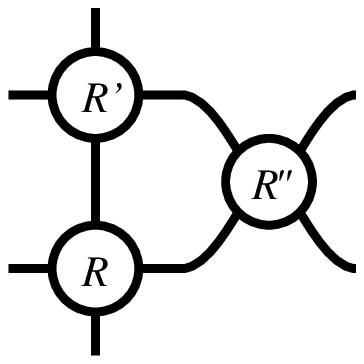}}\;\;=\;\;
\vcenter{\epsffile{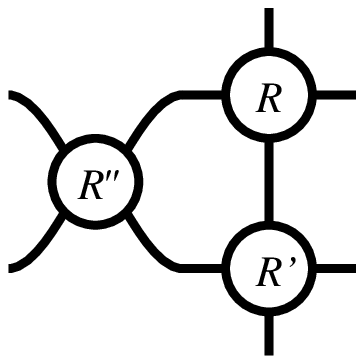}}\; .
\label{ybepic}
\ee
Here
\be
\vcenter{\epsffile{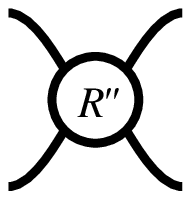}}\;\; = \;\; a'' \;\;\vcenter{\epsffile{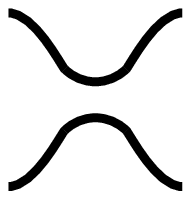}}
\; + b''\;\; \vcenter{\epsffile{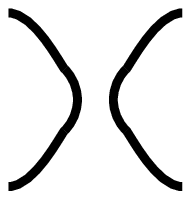}} \; + c''\;\;
\vcenter{\epsffile{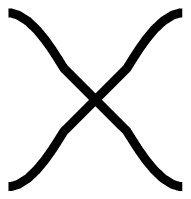}}\; .
\ee

Explicitly, the Yang-Baxter equation reduces to the following set of four
equations
\bea
cb'a'' & = & bb'c'' + bc'a'' \nonumber\\
ac'a'' & = & aa'c''+ca'a'' \nonumber\\
ab'c'' & = & ac'b'' + cb'b'' \nonumber\\
ab'a'' & = & qba'b'' + aa'b''+ca'b'' + bb'b'' +bc'b'' + ba'a'' + ba'c''
% in old convention (not usual):
%ca'b'' & = & aa'c'' + ac'b'' \nonumber\\
%bc'b'' & = & bb'c''+cb'b'' \nonumber\\
%ba'c'' & = & bc'a'' + ca'a'' \nonumber\\
%ba'b'' & = & qab'a'' + bb'a''+cb'a'' + aa'a'' +ac'a'' + ab'b'' + ab'c''
\label{ybe}
\eea
Regarding these equations as a linear system for the unknowns $a''$,
$b''$, $c''$ and requiring the determinant of the  coefficients of the first
three equations to vanish leads to
\be ab(a'+b')c' = a'b'(a+b)c.\label{ybedet}\ee
If we want solutions with positive $a$, $b$ and $c$ we must have
\be \frac{ab}{c(a+b)} = \frac{a'b'}{c'(a'+b')}=\kappa.\ee
The fourth equation from the set~(\ref{ybe}) then implies that
\be \kappa= 1-q/2.\label{defkappa}\ee
We can thus parametrize the mirror weights as follows
\bea
a(u) &=& \rho(u) \cdot (1-u)\nonumber\\
b(u) &=& \rho(u) \cdot u\nonumber\\
c(u) &=& \rho(u) \cdot \kappa u(1-u),
% old convention:
%a(u) &=& \rho(u) \cdot u\nonumber\\
%b(u) &=& \rho(u) \cdot (1-u)\nonumber\\
%c(u) &=& \rho(u) \cdot \kappa u(1-u),
\eea
with $\kappa$ given by eq.~(\ref{defkappa}).
The function $\rho(u)$ is an arbitrary normalization. Furthermore we will
assume that $0 < \kappa \le 1$ so that the model allows for intersections
and has a physical regime $0 < u < 1$ where $a$, $b$ and $c$ are positive.
In the sequel we will take $\rho(u) = 1$, unless otherwise stated.
The spectral parameters satisfy the difference property $u'' = u' -u$.

The proof that a solution of the Yang-Baxter equations leads to commuting
transfer matrices, i.e.\
\be
{\sf T}(u) {\sf T}(v) = {\sf T}(v) {\sf T}(u)\quad\forall u,v
\ee
is the same as usual. It makes use of the periodic boundary conditions,
the existence of a `horizontal'
inverse of the `R-matrix' for generic values of the spectral parameter:
\be
\vcenter{\epsffile{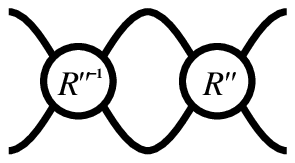}} \;\; =\;\; \vcenter{\epsffile{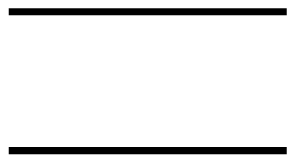}}
\;\; =\;\; \vcenter{\epsffile{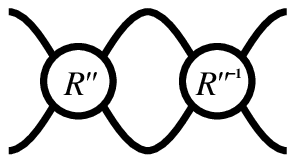}}
\label{inverse}
\ee
and repeatedly uses the YBE:
\bea
{\sf T'} {\sf T} & = & \;\vcenter{\epsffile{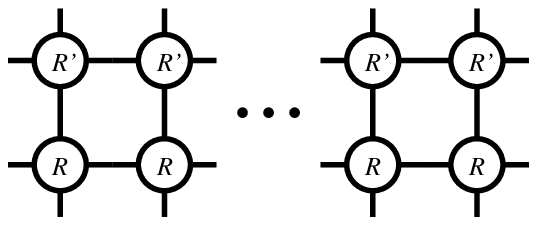}} \nonumber\\
&=&\;\vcenter{\epsffile{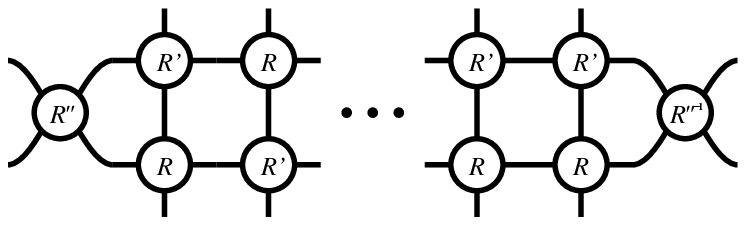}} \nonumber\\
&=&\;\vcenter{\epsffile{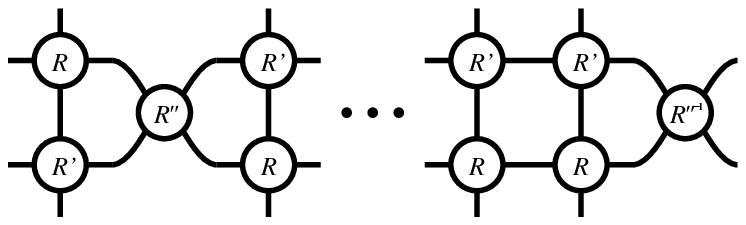}} \nonumber\\
&=&\;\vcenter{\epsffile{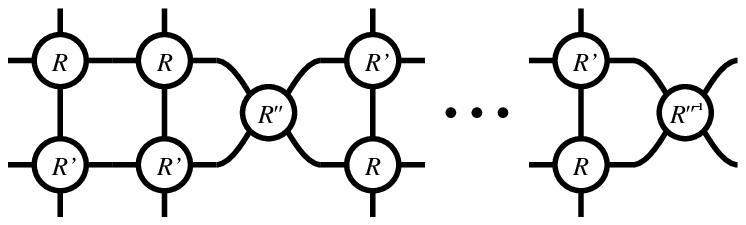}} \nonumber\\
&=&\cdots\nonumber\\
&=&\;\vcenter{\epsffile{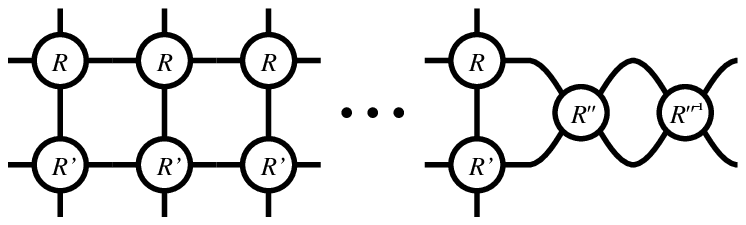}} \nonumber\\
&=&\;\vcenter{\epsffile{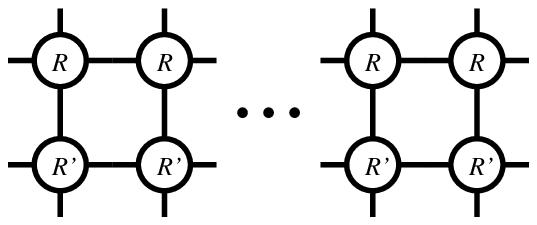}} \nonumber\\
&=&{\sf T} {\sf T'}.\eea

There are four points where the R-matrix $R(u'')$ doesn't have a horizontal
inverse, viz. $u''= \pm 1$ and $u''=\pm \kappa^{-1}$.
At $u''=1$ and $u''=\kappa^{-1}$ it is a multiple of a projector, i.e.\ at
these points there is a non-singular constant $\lambda$ such that the
R-matrix satifies
\bea
R \cdot R &=& \lambda R,\\
\noalign{\hbox{so that}}
\lambda^{-1}R\cdot({\sf id}_2 - \lambda^{-1} R) &=& 0,
\eea
where ${\sf id}_2$ denotes the identity element in the `horizontal' Brauer
algebra $B_2(q)$.
This implies that for these two values of $u''$ the matrix products ${\sf
T}(u+u''){\sf T}(u)$ split in the sum of two terms, one with local
building blocks
\be \lambda^{-1}\;\vcenter{\epsffile{ybel.ps}}\;,\label{firstweight}\ee
the other with building blocks
\be\vcenter{\epsffile{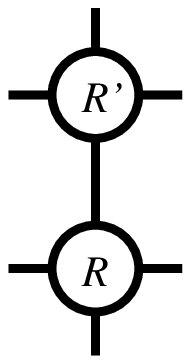}}\;\;\; - \lambda^{-1}\;
\vcenter{\epsffile{ybel.ps}}\;.\label{secondweight}\ee
For $u''= 1$ (and hence $u'=u+1$) the first term~(\ref{firstweight})
enormously simplifies. At this point $\lambda = q$ and the YBE becomes
\be
\vcenter{\epsffile{rpr.ps}}\vcenter{\epsffile{rppb.ps}}\;\;\; = \;\;\;
\vcenter{\epsffile{rppb.ps}}\vcenter{\epsffile{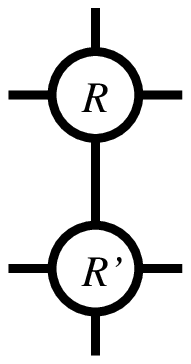}}\;.
\ee
Since on the r.h.s.\ the two points on the left are always connected, the
same must hold for the l.h.s. It now follows that
\be
\vcenter{\epsffile{rpr.ps}}\vcenter{\epsffile{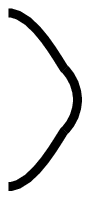}}\;\;\; =
\phi(u) \;\;\vcenter{\epsffile{lhalfrppb.ps}}
\;\;\;\vcenter{\epsffile{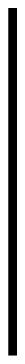}}\;.
\ee
It is easy to see that
\be\phi(u) = ab'  + cc' = (1-u^2)(1-\kappa^2u^2).\ee
Furthermore,
\be
q^{-1} \vcenter{\epsffile{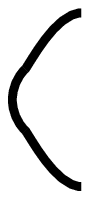}}\vcenter{\epsffile{lhalfrppb.ps}}
\;\; = 1 \ee
and the second set of configurations~(\ref{secondweight})
turns out to have an overall factor $u$.
We thus obtain the inversion identity
\be \label{invid}
{\sf T}(u+1) {\sf T}(u) = \phi(u)^N {\sf id} + u^N {\sf T}^{(2)}(u),
\ee
where the `fused' transfer matrix ${\sf T}^{(2)}(u)$ is a polynomial of
degree $3N$ in $u$ over the Brauer algebra $B_N(q)$.

As for many other models, for large $N$ and small $u$ the first term on the
r.h.s.\ of eq.~(\ref{invid}) dominates the second, hence every eigenvalue
$\Lambda(u)$ of ${\sf T}$ satisfies
\be \label{invrel}
\psi(u+1) \psi(u) = \phi(u),
\ee
where
\be
\psi(u) =  \lim_{N\to\infty} \left(\Lambda(u)\right)^{1/N}.
\ee
In particular the `partition function per site' $\psi_0(u)$ satisfies
the inversion relation~(\ref{invrel}). It is fixed by the symmetry $u\to
1-u$ (related to an involutive automorphism of the Brauer algebra) to be a
product
\be
\psi_0(u) = F(u) F(1-u)
\ee
and the function $F(u)$ satisfies
\be
F(1+u) F(u) = (1+u) (1+\kappa u).
\ee
The solution that is free of zeroes in the `physical' strip $0< {\rm Re\,}u
<1$ is
\be
F(u) = 2\sqrt{\kappa}\, \frac{\Gamma(1 + \frac{u}{2})}
{\Gamma(\frac{1}{2} + \frac{u}{2})} \frac{\Gamma(\frac{1}{2} +
\frac{1}{2\kappa}+\frac{u}{2})}{\Gamma(\frac{1}{2\kappa} + \frac{u}{2})}.
\ee
Hence the result for the normalized partition function per site (with
$a+b+c=1$) is
\be
\psi_{0\rm n} = \frac{4\kappa}{1+\kappa u(1-u)} \frac{\Gamma(1 +
\frac{u}{2})}{\Gamma(\frac{1}{2} + \frac{u}{2})}
\frac{\Gamma(\frac{3}{2}-\frac{u}{2})}{\Gamma(1-\frac{u}{2})}
\frac{\Gamma(\frac{1}{2} + \frac{1}{2\kappa} + \frac{u}{2})}
{\Gamma(\frac{1}{2\kappa}+\frac{u}{2})}
\frac{\Gamma(1+\frac{1}{2\kappa}-\frac{u}{2})}{\Gamma(\frac{1}{2}+
\frac{1}{2\kappa}-\frac{u}{2})}.
\ee
When we take the logarithmic derivative of the normalized partition function
with respect to $\kappa$, keeping $u$ fixed, we get the linear combination
of densities of mirrors and loops:
\be
\frac{\partial}{\partial\kappa} \log\psi_{0\rm n} =
\frac{\partial a}{\partial\kappa} \frac{\langle\rho_a\rangle}{a}
+\frac{\partial b}{\partial\kappa} \frac{\langle\rho_b\rangle}{b}
+\frac{\partial c}{\partial\kappa} \frac{\langle\rho_c\rangle}{c}
+\frac{\partial q}{\partial\kappa} \frac{\langle\rho_{\rm loops}\rangle}{q}.
\label{dpsidk}\ee
When $q=1$ ($\kappa = \half$) the densities of the various types of mirrors
are are simply equal to their normalized Boltzmann weights, since there is
no coupling between different mirrors (the partition function for $q=1$ is
$Z=(a+b+c)^{\#\rm sites}$). Then eq.~(\ref{dpsidk}) reduces to
\be
\langle\rho_{\rm loops}\rangle_{q=1} = -\frac{1}{2}\left.
\frac{\partial}{\partial\kappa}\log\psi_{0\rm n}\right|_{\kappa=1/2}.
\ee
At the isotropic point $u=\half$, which corresponds to the mirror weights
\be
a=b=\frac{4}{9},\quad c=\frac{1}{9}, \label{specialvalues}
\ee
the loop density is
\bea
\langle\rho_{\rm loops}\rangle_{q=1} &=& \textstyle
2\left( \psi_2(\frac{7}{4}) - \psi_2(\frac{5}{4})\right) -
\frac{8}{9}\nonumber\\
&=&\textstyle 2\pi - \frac{56}{9},
\eea
with $\psi_2$ the digamma function. This number can be measured in a Monte
Carlo simulation, by setting the mirrors according to the probability
distribution eq.~(\ref{specialvalues}) and counting the number of paths. The
numerical result agrees with the exact answer to five
significant digits~\cite{BBKM}.

The loop density also gives information about the length distribution of
loops passing through a given edge: the average inverse length is given by
\be
\langle L^{-1}\rangle = \half\langle\rho_{\rm loops}\rangle_{q=1}.
\ee

A more extensive and detailed study of this loop model will be published
elsewhere~\cite{rietman+nienhuis}.
\leftline{{\bf Acknowledgements}}
\noindent We thank Marco Brummelhuis, Dirk Jan Bukman, Hans Kuijf and Steven
van der Marck for discussions in the early stages of this work and for
performing the Monte Carlo simulations.
We also thank Paul Pearce and Vladimir Bazhanov for discussions and some
useful remarks.\par
\noindent This work is financially supported by the Stichting voor
Fundamenteel Onderzoek der Materie.


\begin{thebibliography}{99}
\bibitem{cohen} T. W. Ruijgrok and E. G. D. Cohen {\sl Phys.\ Lett.\ }{\bf A
133} (1988) 415; X.~P.~Kong and E.~G.~D.~Cohen {\sl Phys. Rev.\ }{\bf B 40}
(1989) 4838,  {\sl J.~Stat.\ Phys. }{\bf 62} (1991) 1153; R.~M.~Ziff,
X.~P.~Kong and E.~G.~D.~Cohen {\sl Phys.\ Rev.\ }{\bf B 44} (1991) 2410.
\bibitem{blotnight} H. W. J. Bl\"ote and M. P. Nightingale {\sl Physica\/ }
{\bf 112 A} (1982), 405.
\bibitem{brauer} R. Brauer {\sl Ann.\ Math.\ }{\bf 38} (1937), 857.
\bibitem{BBKM} M. Brummelhuis, D.~J.~Bukman, H.~Kuijf and S.~van~der~Marck,
private communication.
\bibitem{rietman+nienhuis} R. Rietman and B. Nienhuis, in preparation.
\end{thebibliography}
\end{document}